\documentclass[pra,twocolumn]{revtex4-1}

\usepackage{graphicx}
\usepackage{amsmath}
\usepackage[latin1]{inputenc}
\usepackage{tikz}
\usepackage{float}
\usepackage{color}
\usepackage{graphicx}
\usepackage{epstopdf}
\usepackage{pgfplotstable}
\usepackage{filecontents}
\usepackage{pgfplots} 
\usepackage{hyperref}
\usetikzlibrary{shapes,arrows}
\usepackage{todonotes}

\begin{document}

\title{Coherent transport by adiabatic passage on atom chips}

\author{T.~Morgan$^{1,2}$, L.~J.~O'Riordan$^{1,2}$, N.~Crowley$^1$, B.~O'Sullivan$^1$ and Th.~Busch$^{1,2}$}
\affiliation{$^1$Department of Physics, University College Cork, Cork, Ireland}
\affiliation{$^2$Quantum Systems Unit, OIST Graduate University, Okinawa, Japan}

\begin{abstract}
Adiabatic techniques offer some of the most promising tools to achieve high-fidelity control of the centre-of-mass degree of freedom of single atoms. As their main requirement is to follow an eigenstate of the system, constraints on timing and field strength stability are usually low, especially for trapped systems. In this paper we present a detailed example of a technique to adiabatically transport a single atom between different waveguides on an atom chip. To ensure that all conditions are fulfilled, we carry out fully three dimensional simulations of the system, using experimentally realistic parameters. We also detail our method for simulating the system in very reasonable timescales on a consumer desktop machine by leveraging the power of GPU computing.
\end{abstract}



\maketitle
\section{Introduction}
\label{sec:Introduction}
Recent experimental progress in trapping and controlling all degrees of freedom of single atoms and ions has allowed us to test and
explore the fundamentals of quantum mechanics at a completely new level \cite{Chen:11,Bergmann:98}. In fact, progress has been so dramatic that application of the laws of single and few particle quantum mechanics to areas such as quantum information and quantum metrology has come into experimental reach \cite{Nielsen:00,Riedel:10}.

While control over the internal degrees of freedom of atoms is a highly advanced field, significant progress in developing techniques to coherently control the external degrees of freedom to the same level has only recently been achieved. One class of techniques that can offer high fidelities are adiabatic processes and recently a technique called Coherent Tunnelling by Adiabatic Passage (CTAP) was shown to be a very promising tool for controlling the quantised centre-of-mass state of a single particle trapped in a microtrap \cite{Eckert:04}. CTAP is designed to transfer populations between microtraps at high fidelities while being robust to variations in the system parameters. Although the physics of CTAP is well understood, the process has yet to be observed experimentally and several realistic systems have recently been proposed \cite{Eckert:06,Morgan:11,Kohler:13}.

Coherent transport between microtraps can be facilitated via tunnelling and the tunnelling rates  can be controlled by moving the centres of the individual traps relative to each other. While this requires dynamical potentials, a similar system with static potentials can be constructed by considering three parallel running waveguides with spatially varying coupling strength between them  and an atom which travels along these guides \cite{Eckert:06}. Recently, in our previous work, a realistic atom chip system of this kind was considered \cite{OSullivan:10}, however the simulations were limited to two dimensions.

While the transversal dynamics in a system of waveguides can be well described in a two-dimensional model, effects stemming from bending, longitudinal dispersion and the lack of stationary states in the z-direction cannot be accounted for. To overcome these limitations and understand the total dynamics of a waveguide system, it is necessary to carry out a fully three dimensional simulations.
  
We therefore present here, an analysis of a system composed of three waveguides by taking the full dynamics in all three spatial directions into account and using realistic experimental parameters. The latter is important as most treatments of the problem in recent years have assumed idealized trapping potentials that guarantee resonance between the individual traps at any moment in time. By carrying out three dimensional simulations which account for all possible dynamics, we show that CTAP is indeed a suitable technique for use in waveguides on atom chips.
 
By today, fully three dimensional simulations of the Schr\"odinger equation in the context of atomic transport are still rare \cite{Rab:08}. The computational resources needed are very large and have traditionally required the power of  large supercomputers. Recently it was shown that the emerging technique of GPU (graphics processing unit) computing allows tremendous speedup of many numerical techniques including the fast Fourier transform (FFT) \cite{Bauke:11}, which is the main numerical tool that we require. By making use of this, we have been able to perform the simulations of this extensive atomic system with one consumer desktop PC using the CUDA programming model and numerical libraries, on very reasonable timescales.

The structure of this paper is as follow. In Sec.~\ref{sec:CTAP} we will briefly review the CTAP process in waveguide systems and in Sec.~\ref{sec:Atom Chips} we describe the atom chip potentials we are simulating. In Sec.~\ref{sec:MPICUDA} we will discuss our
implementation of CUDA and MPI (Message Passing Interface) codes and examine the performance benefits in each case.  Our results of the three dimensional simulations and the evidence that CTAP can be observed will be presented and discussed in Sec.~\ref{sec:Results}. Finally we conclude in Sec.~\ref{sec:Conclusions}.

\section{Coherent Tunnelling by Adiabatic Passage}
\label{sec:CTAP}

\begin{figure}[tb]
  \includegraphics[width=\linewidth]{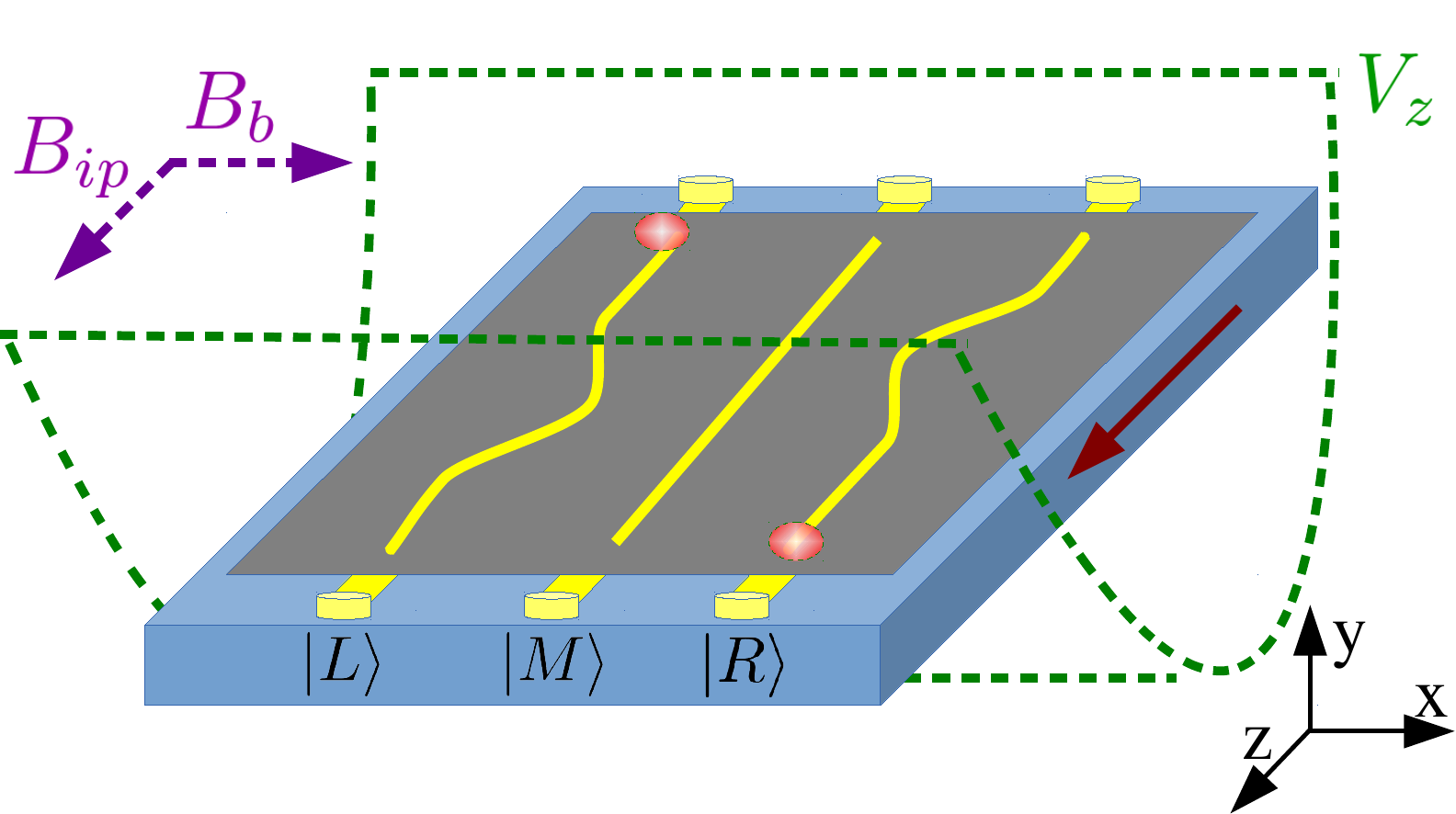}
  \caption{(Color online) Schematic of the suggested setup for observing the CTAP process in a system of waveguides. Note that the asymmetric approach of the outer wires to the middle wire is exaggerated, so that the counter-intuitive arrangement is visible. The atom is initially located in the left guide and, due to the presence of a harmonic oscillator potential $V_{z}$ in the $z$-direction, travels along the direction indicated by the red solid arrow. We also show the expected position of the atom at $t=\pi/{\omega_z}$ in the right hand side guide and indicate the orientation of the bias field, $B_b$, and the applied field, $B_{ip}$ (purple dashed arrows).}
  \label{fig:Schematic}
\end{figure}

Let us first briefly review the CTAP process by considering an atom trapped in a linear system of three
identical, one-dimensional microtraps \cite{Eckert:04}.  Assuming that the atom is in its centre-of-mass ground state in the trap on the left hand side,
$|L\rangle$, it can reach the ground states of the other two traps, $|M\rangle$ and $|R\rangle$, through coherent tunnelling described by the strength $J_{LM}$ for the transition $\vert L \rangle\to\vert M \rangle$ and $J_{MR}$ for $\vert M \rangle\to\vert R
\rangle$. In this basis the Hamiltonian is given by
\begin{equation}
  \label{eq:wgHamiltonian}
  H(t)=\hbar\begin{pmatrix}
                    0             & -J_{LM}(t) & 0  \\
                   -J_{LM}(t)  & 0            & -J_{MR}(t) \\
                    0             & -J_{MR}(t) & 0  
  \end{pmatrix} ,
\end{equation}
where the energy of the trap ground states was re-normalized to zero. The tunnelling strengths are assumed to be time-dependent, which can be achieved by increasing or decreasing the distances between neighbouring traps, $d_{LM}(t)$ and $d_{MR}(t)$. The eigenstates of the Hamiltonian \eqref{eq:wgHamiltonian} are well known \cite{Bergmann:98} and of particular interest for adiabatic transport is the so-called dark state
\begin{equation}
  |d\rangle=\cos\theta|L\rangle-\sin\theta|R\rangle,
\end{equation}
in which the  mixing angle $\theta$ is given as a function of the tunnelling strengths as
\begin{equation}
  \tan\theta=J_{LM}/J_{MR}.
\end{equation}
This state has a non-degenerate zero eigenvalue and an adiabatic evolution will therefore guarantee that the system, once prepared in
$|d\rangle$, will always stay in it. Note that the only contribution of $|M\rangle$ to $|d\rangle$ is through the mixing angle and that the system has zero probability to be found in $|M\rangle$ at any time.

The CTAP process can now be understood by considering an atom initially in the state $|L\rangle$. Increasing and decreasing $J_{MR}$
before $J_{LM}$, which is counter-intuitive to traditional tunnelling schemes, continuously decreases the population in state $|L\rangle$
and increases the population in state $|R\rangle$, leading to a 100\% transfer at the end of the process.

Adapting this process to a system of waveguides is now straightforward. The temporal dependence of the tunnelling strength in eq.~\eqref{eq:wgHamiltonian} can be replaced by a spatial one through suitable adjustment of the distance between neighbouring waveguides as a function of the direction the particle travels in (see Fig.~\ref{fig:Schematic} for a schematic view) \cite{Eckert:06}.

There are, however, several conditions that both, the microtrap and the waveguide system, must fulfil for the CTAP dynamics to occur. Firstly, the process must be adiabatic with respect to the other relevant energy scales in the system. For the waveguide system this means the whole process has to be slower than the inverse of the approximate transverse trapping frequencies of the guides. As typical numbers for such guides are in the kHz regime, this means that the time allowed for the atom to travel along the chip  can be much shorter than a typical system's lifetime. The second condition which has to be fulfilled, as previously mentioned, is that all trapping states are in resonance at any point in time, which is difficult to achieve once the potentials of the individual guides start to overlap. However, we will demonstrate in the next section how a waveguide setup on an atom chip is a realistic experimental system in which this resonance condition can be fulfilled to a good approximation.

\section{Atom Chips}
\label{sec:Atom Chips}

Atom chips are versatile experimental tools that are by today used extensively in experiments with ultra-cold atoms \cite{Folman:00,Fortagh:07}. A small current flowing through nano-fabricated wires on the substrate produces a magnetic field gradient in such a way, that cold atoms can be trapped very close to the surface. Because the layout of the nanowires can be chosen during the chip's production process, atom chips have been used in many cold atom experiments to produce microtraps, interferometers and waveguides \cite{Fortagh:07,Schwindt:05,Schumm:05,Petrovic:11}. Here we will take advantage of this versatility to consider waveguides in the geometry indicated in Fig.~\ref{fig:Schematic} and develop a procedure which will allow to observe high fidelity transport based on CTAP.

Let us briefly review the basic description and properties of atom chip trapping. The magnetic potential $\bf B$ at position $\bf r$ generated by a typical nanowire on an atom chip can be described by the Biot-Savart law
\begin{equation}
   {\bf B}= \frac{\mu_0 I}{4\pi} \oint \! \frac{d{\bf l}  \times {\bf{\hat r}}}{{r^2 }},
\end{equation}
where $I$ is the current in the wire, $\mu_0$ the vacuum permeability, $\bf{\hat r}$ the unit vector in the direction of $\bf r$ and $d{\bf l}$ is the differential length of the wire carrying current $I$. For this expression to be valid, however, we have to assume that the thickness of the wires is negligible, which is a good approximation as long as we are using the properties of the field at a sufficient distance above the chip's surface. To achieve this and to lift the field minima above the nanowires for the desired waveguide structure, a homogeneous magnetic bias field, $B_b$, can be applied orthogonal to the current flow. This raises the potential minimum to a height above the wire given by
\begin{equation}
  r_0=   \frac{\mu_0}{2 \pi} \frac{I}{B_b}.
\end{equation}
Finally, to lift the degeneracy of the spin states of the atoms and avoid losses due to spin flips at the centre of the waveguide a further magnetic field, $B_{ip}$, parallel to the direction of the wires is usually applied.

\begin{figure}
  \includegraphics[width=\linewidth]{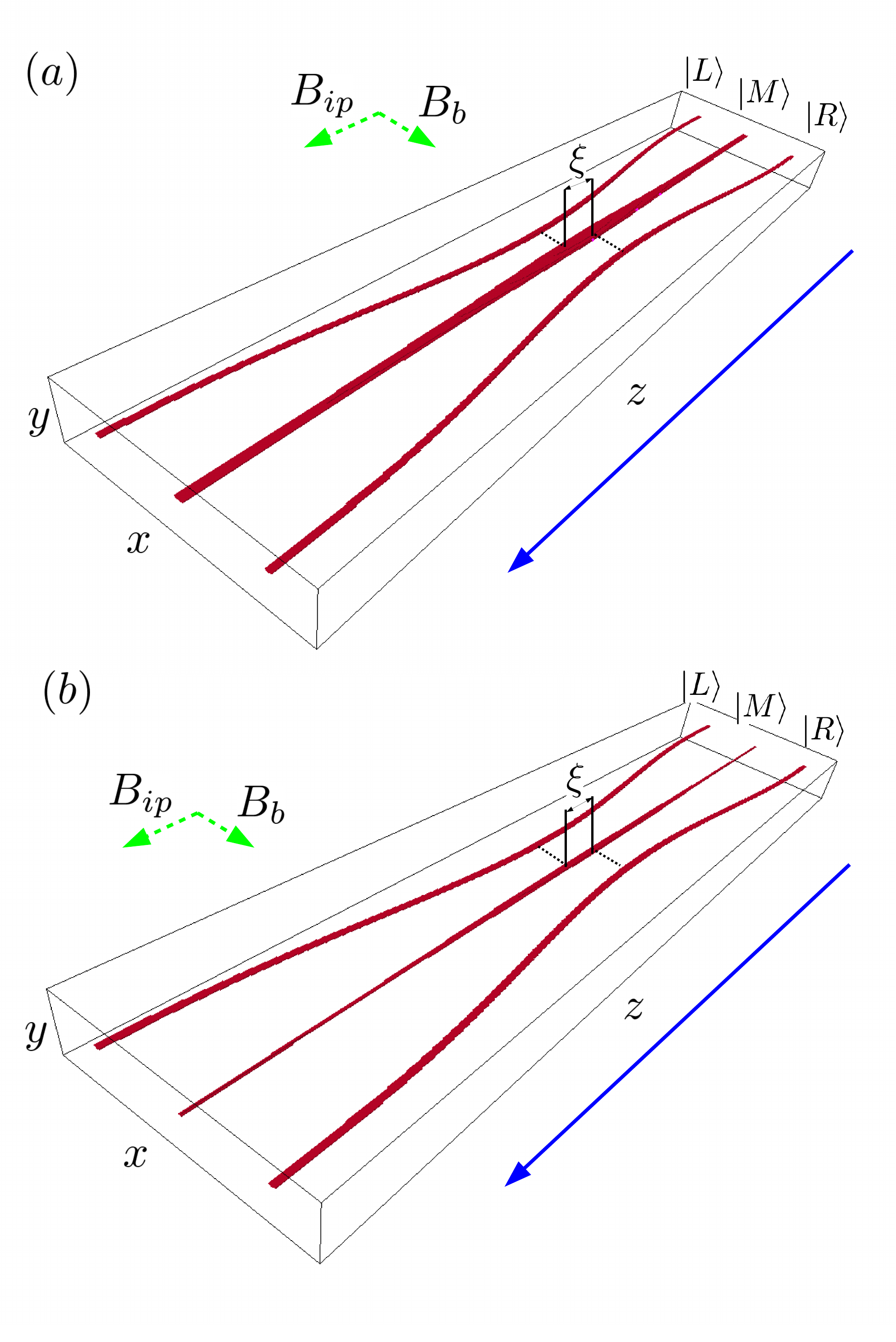}
  \caption{(Color online) Isosurfaces of the waveguides created on an atom chip with the direction of propagation indicated by the blue solid arrow (for clarity $V_z=0$ in this plot). The dimensions of the interesting area on the chip we simulate are $20\mu$m $\times 1000\mu$m ($x \times z$) and we take a height ($y$ direction) above the chip of $4\mu$m into account. The three wires are initially equally separated by $7\mu$m and their distance at the position of closest approach is $4.3\mu$m. The left wire remains straight initially for a distance of $50\mu$m, which produces an asymmetry in the point of closest approach of the left and right wires to the middle wire as indicated by $\xi$. The bias and applied fields (indicated by the green dashed arrows) are $B_b=140 \times 10^{-4}$T and $B_{ip}=300 \times 10^{-4} $T. In (a) the currents of the left, middle and right wires are $I_L=I_M=I_R=0.1$A respectively and in (b) the currents of the left and right wires are $I_L=I_R=0.1$A and the middle wire current is reduced to $I_M=0.07$A.}
      \label{fig:Potentials}
\end{figure}

\begin{figure}
  \includegraphics[width=\linewidth]{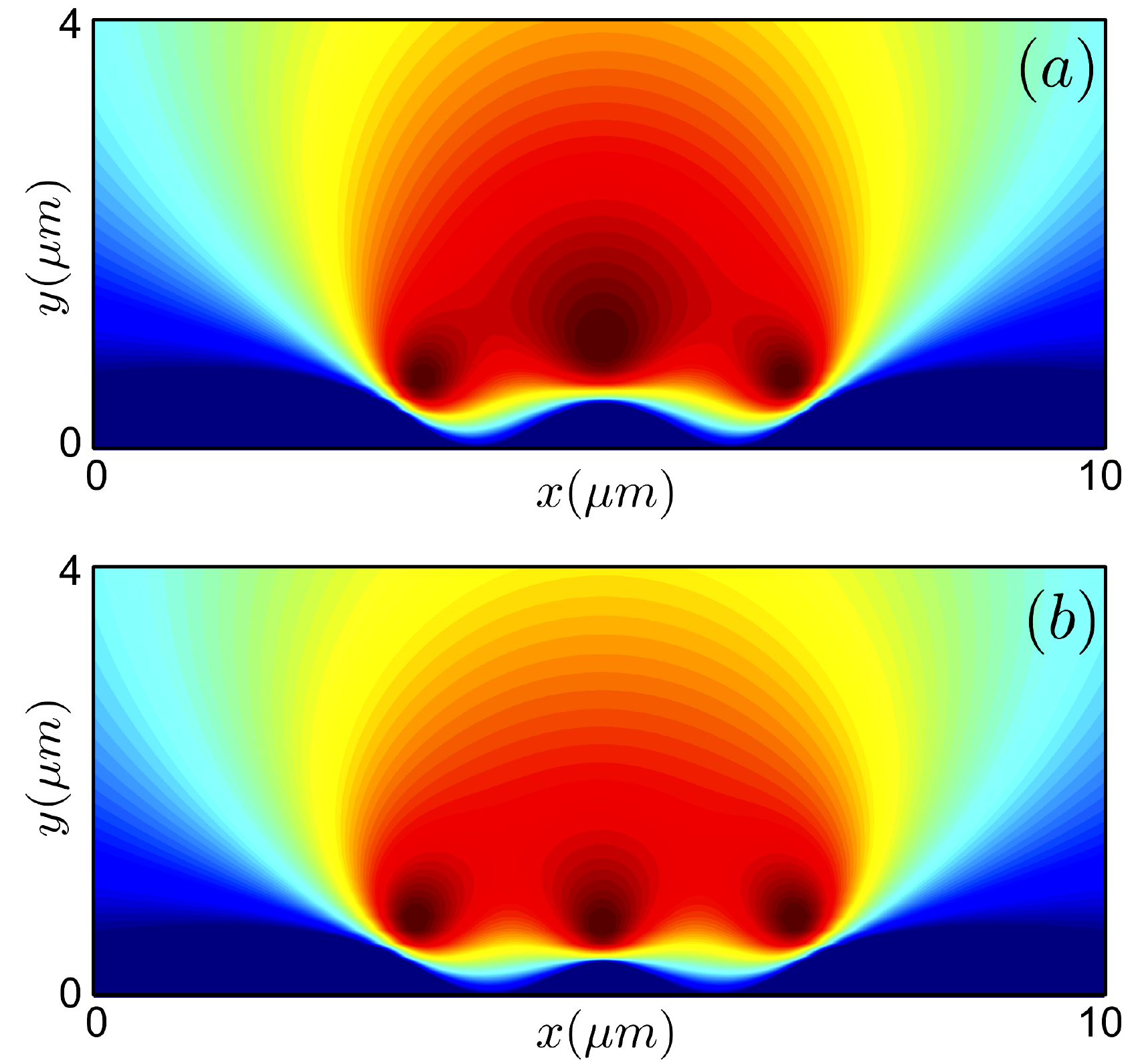}
  \caption{(Color online) Contour plot of the waveguides at 500$\mu m$ along the z-axis. Panel (a) shows the deformation of the waveguides when all currents are equal, $I_L=I_M=I_R=0.1$A and panel (b) shows how this effect can be mitigated by using a reduced middle wire current of $I_M=0.07$A, while the current in the outer wires remain at $I_L=I_R=0.1$A.}
  \label{fig:slice}
\end{figure}

An example of the waveguide potentials resulting from this model for $^6$Li  atoms and for experimentally realistic parameters is shown in Fig.~\ref{fig:Potentials}. If an atom is initially located in the left waveguide and travels in the positive $z$-direction, these waveguides provide the desired counter-intuitive tunnel coupling needed for CTAP. To give the atom momentum to travel along the wires we add an additional harmonic oscillator potential, $V_z$, of frequency $\omega_z$ along the $z$-direction, which is centred at the middle of the chip (see Fig.~\ref{fig:Schematic}). This will also lead to a re-focusing of the travelling wavepacket at the classical turning point on the other side of the chip and help to clearly determine the position of the atom.

To ensure that the process is adiabatic and the atom remains in the dark state of the system at all times, the total time for the process has to be much larger than the inverse of the transverse trapping frequencies of the individual waveguides. By approximating the potentials to have a harmonic oscillator shape in the transverse direction, we find the inverse of the relevant frequency to be of the order of $f_{HO} ^{-1}\approx 0.2$ ms and by choosing the trapping frequency of the harmonic oscillator in the z-direction to be $\omega_z=2 \pi \times 5$ Hz, the total time taken for the process (half an oscillation) is $0.1$ s. This allows to clearly fulfil the adiabaticity condition at any point during the evolution. 

Finally, the bend in the wires will lead to a potential from the currents in the $z$-direction, which requires the atom to have enough kinetic energy to overcome it and therefore sets an upper bound to the adiabaticity that can be reached. However, this effect can be reduced by increasing the length of the atom chip (z-direction) and therefore reducing the curvature of the wires. From our simulations, we find that the kinetic energy resulting from locating an atom initially at the edge of a chip that is $z_\text{max}=1000 \mu$m long allows us to successfully propagate the atom though the waveguides using the harmonic trap described above.

\section{MPI and CUDA}
\label{sec:MPICUDA}

To simulate the propagation of the atom along the waveguide we solve the three-dimensional time-dependent Schr\"odinger equation using the well known Fourier Split Operator method \cite{Fleck:splitop}. A typical numerical implementation of this method requires the use of 4 Fourier transforms followed by 3 complex multiplications for each time step. The numerical library we make use of to perform the Fourier transforms is the well known FFTW library, and its GPU implementation CUFFT \cite{Nvidia:cufft_4_1}.

Performing three dimensional Fourier transforms is the most intensive part of our code with the length of time required to perform one iteration of the split operator method depending heavily upon the size of the numerical grid. As discussed in the previous section, the atom chip has a relatively large extension in the $z$-direction  ($z_\text{max}=1000 \mu$m) compared to the other dimensions. Since the maximum value of the momentum grid is defined as $p_\text{max} = \frac{\pi N_z}{z_\text{max}}$ we require a large number of points, $N_z$, for our grid to be large enough to resolve the longitudinal momentum stemming from the external harmonic oscillator potential. This is the main reason that the computational resources required to simulate the system are quite substantial.

\subsection{GPU Computing}

To overcome the numerical barrier presented by this system we turn to the relatively new computing paradigm of GPU (graphics processing unit) computing. Whereas traditional computers perform computations using the central processing unit (CPU), GPU computing allows some of the work to be offloaded to the graphics processor. GPUs are inherently SIMD (single instruction, multiple data) devices, designed for operating upon a large data set at a given time with a single task, such as a 2D grid of pixels. Due to their parallel nature, GPUs can perform better than CPUs for certain types of calculations. One example where they offer large performance gains are fast Fourier transformations (FFTs) and it was recently shown that the Fourier split operator method can be accelerated using GPU computation \cite{Bauke:11}. This performance increase offers the numerical power needed to simulate the above system and we have implemented the algorithms for split-operator evolution of the Schr\"odinger equation with C, CUDA and Nvidia's CUFFT libraries for the Fourier transforms. 

\subsection{Performance}
To demonstrate the performance offered by GPU computing we compare it to using FFTW with MPI, a more traditional CPU-based method. The message passing interface (MPI) implementation allows the code to be run across multiple machines, benefiting from the parallelism which may be offered by a supercomputing cluster. Although MPI-enabled FFTW is fast and supports extremely large grid sizes, it requires computer-cluster access of a significant size to be a viable option. 

To effectively simulate the CTAP process and accurately resolve the momentum, our code requires a grid size of $256\times 64\times1024$ ($x\times y\times z$). For accurate time evolution a timestep of $\Delta t = 1\times 10^{-6}$ s was found to be adequate. For the GPU simulations, the test system was an Intel Core i7 2600K CPU at stock frequency, 8GB DDR3 memory operating at 1600 MHz, 7200 RPM HDD, Nvidia GeForce GTX 580 with 3GB of onboard memory running at 783 MHz GPU core frequency, 1566 MHz shader processor frequency, and 2010 MHz memory frequency. For all simulations the desktop was running Ubuntu 11.10 64-bit operating system and all calculations were performed in double precision (64-bit floating point) where applicable. For the CPU simulations we utilized the supercomputers at ICHEC (Irish Centre of High-End Computing).

Table \ref{tbl:timing} shows the approximate timings for the completion of runs on GPU and CPU. As one can see, not only does GPU computing offer a 6-fold improvement over a single CPU, it also allows us to achieve a performance level which is comparable to a 64 core CPU. Such performance has previously been restricted to high powered supercomputers. Having such computational power available to a single user on a Desktop computer allows us to obtain a large volume of simulated data on a much shorter timescale rather than through the use of a shared resource CPU-based computer cluster. Additionally, a second GPU card added to the system allowed concurrent runs of the code, which effectively halved the overall time required for a large number of simulations. It is also worth mentioning that moving computations over to the GPU of the system frees up the CPU and a large part of the system memory to be used for other task that would have previously been inhibited by CPU bound computations.

\begin{table}[tb]
  \begin{center}
    \begin{tabular}{|c||c|c|c|}
      \hline
      Device & Num. Devices & Timing  & Rel. Improvement \\ \hline
      CPU (MPI) & 8 & $\sim$6 Hr & 1.0$\times$ \\ 
      & 16 & $\sim$4 Hr & 1.5$\times$ \\
      & 32 & $\sim$1.5 Hr & 4.0$\times$ \\ 
      & 64 & $\sim$1 Hr & 6.0$\times$ \\ \hline
      GPU & 1 & $\sim$1 Hr & 6.0$\times$ \\ \hline
    \end{tabular}
  \end{center}
   \caption{The approximate times taken to simulate the propagation of an atom through our atom chip system on both GPU and CPU.}
   \label{tbl:timing}
\end{table}

\section{3D Simulations}
\label{sec:Results}
In the following section we will present a set of typical results from the GPU accelerated 3D simulations we have carried out over a large range of experimentally controllable parameters and show that the atom chip allows for the CTAP process to take place. All parameters for our atom chip are the same as in Fig.~\ref{fig:Potentials} unless otherwise stated.

Our simulations start out with a single $^{6}$Li atom which is initially located in the left waveguide. Its transversal wavefunction corresponds to the ground state of the potential in the transversal direction (determined numerically) and longitudinally we assume a Gaussian profile of similar width. We then evolve this initial state in time and due to the longitudinal harmonic oscillator potential centred at the middle of the atom chip ($z=500 \mu$m), the atom starts to propagate along the waveguide. 

Initially the wires are far enough away from each other for each waveguide to be approximately given by the current of the wire closest to it and if all currents are identical, the waveguides are in resonance. However, once the wires start approaching each other, the respective magnetic fields add and create waveguide potentials of unequal size (see Figs.~\ref{fig:Potentials} (a) and \ref{fig:slice} (a)). This drives the transversal ground states of the guides out of resonance and the conditions for observing the CTAP process are no longer given.
 
However, atom chips offer an intriguingly straightforward way to adjust for this, as the current in each wire can be individually (and even time-dependently) controlled. This can be used to compensate for effects stemming from the potentials overlapping and ensure resonance between the waveguides \cite{OSullivan:10}. While one can imagine a numerically optimised algorithm that adjusts the currents in a time-dependent manner based on the position of the centre-of-mass of the atom, here we will show that a much simpler approach, which maintains the simplicity of all currents being constant in time, is already sufficient. We suggest to reduce the current in the middle wire so that in the crucial coupling region, where the magnetic fields from neighbouring waveguides have the strongest influence on each other, the waveguides are approximately resonant. 

To demonstrate the effect of this adjustment we show in Fig.~\ref{fig:slice} a transversal cut through the system at the middle of the chip ($z=500 \mu$m) for the case where (a) all three currents are identical ($I=0.1$A) and (b) where the current in the middle wire is reduced ($I_M=0.07$A). One can clearly see that the transversal shape of the waveguides is very similar for the case of the reduced centre current, which indicates that this approach can lead to enhanced resonance between the guides.

In the areas where the guides are further away from each other, however, the reduced current in the middle wire will have the opposite effect and reduce the quality of the resonance. This can clearly be seen from the iso-potential surface plot in Fig.~\ref{fig:Potentials} (b). Yet, since the tunnelling in these areas is small, it has only a negligible influence on the CTAP process and we will in the following demonstrate that the near resonant setup of Fig.~\ref{fig:slice} (b) allows to observe the CTAP process. 

\begin{figure}[tb]
  \includegraphics[width=\linewidth]{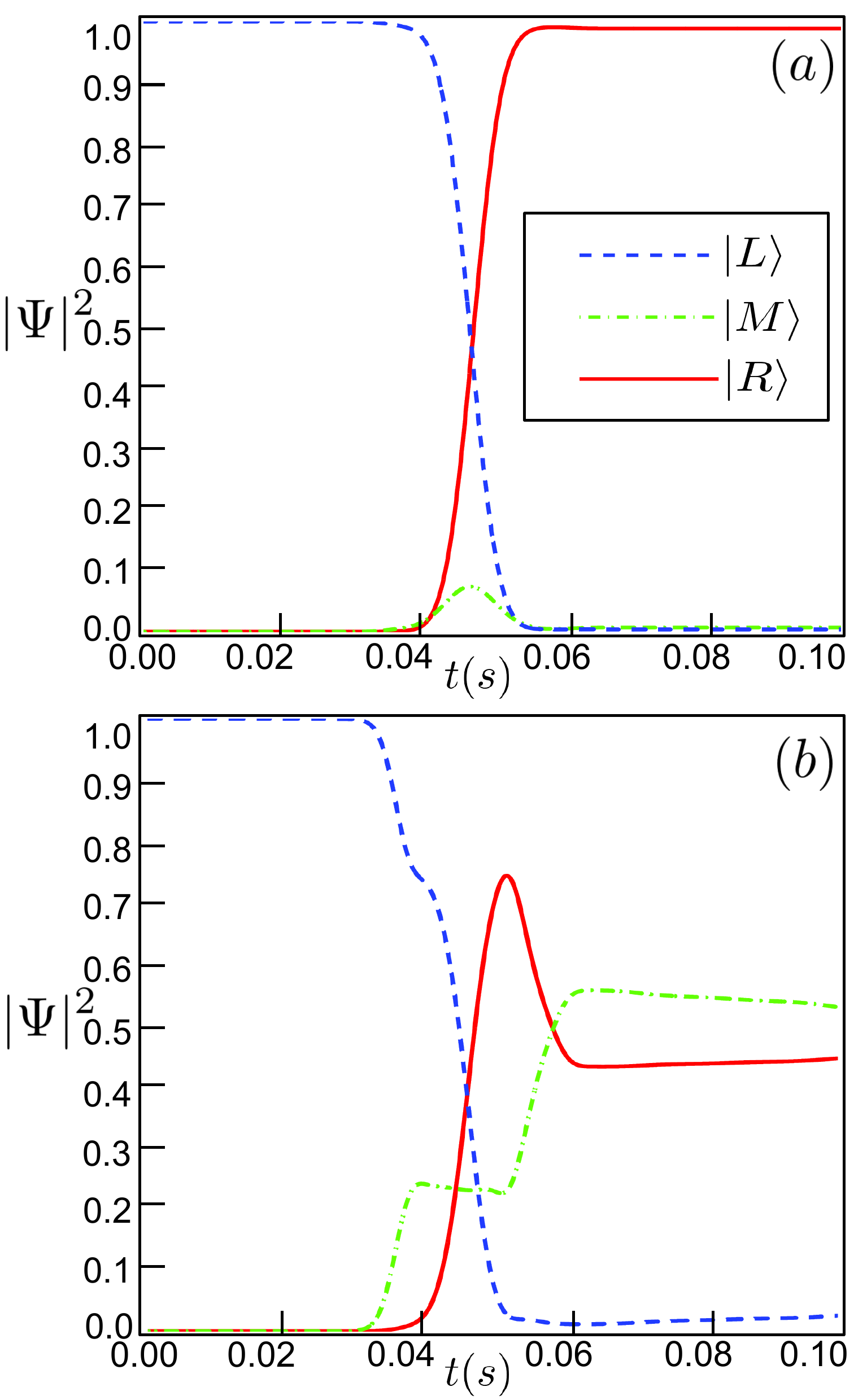}
  \caption{(Color online) The population in the left (blue dashed line), middle (green dot-dash line) and right (red solid line) waveguides as a function of time for (a) the counter-intuitive waveguide arrangement and (b) for intuitive, direct tunnelling one. The current in the middle wire is reduced to $I_M=0.07 A$.}
  \label{fig:Populations}
\end{figure}

In Fig.~\ref{fig:Populations} we show the population in each waveguide as a function of time for an atom chip with reduced current in the central wire. The results in Fig.~\ref{fig:Populations} (a) are obtained for the situations where the wires are arranged such that the counterintuitive tunnelling sequence takes place and full transfer from the initial guide into the final guide is clearly visible. Only a small population in the central guide appears halfway through the process, and while the ideal CTAP process does not allow for population in the central trap at any time, the limited adiabaticity and resonance of the simulated setup leads to this temporary deviation. However, it has no effect on the final state.

In contrast to this, and confirming that the large fidelity of the transport process above is due to CTAP, we show in Fig.~\ref{fig:Populations} (b) the results for an intuitive arrangement of the waveguides on the atom chip. As is clearly visible, this does not produce high fidelity population transfer to the guide on the right hand side, but rather leads to a split of the probability between the middle and the right hand side wire. 

\begin{figure}[tb]
  \includegraphics[width=\linewidth]{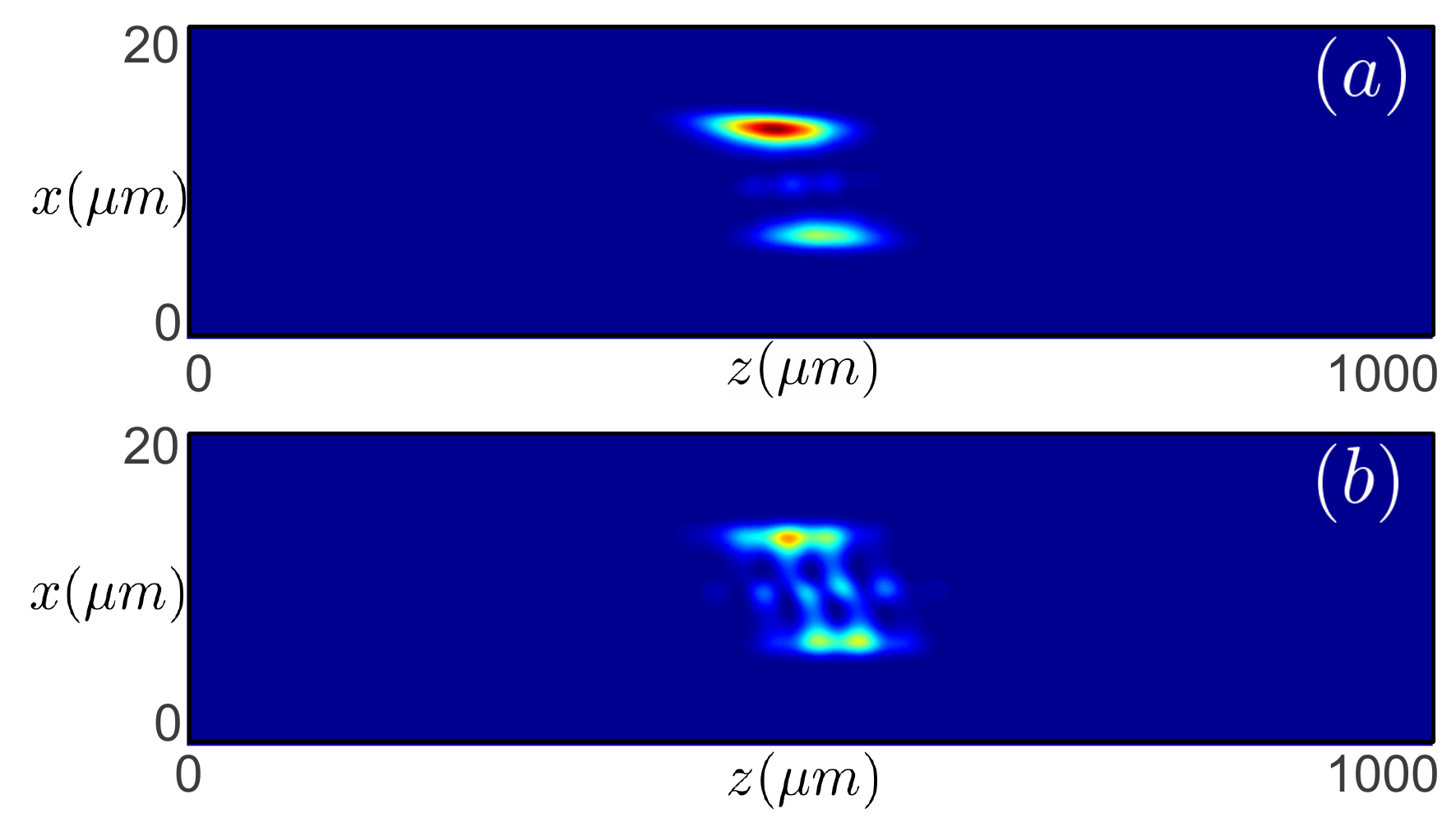}
  \caption{(Color online) The density of the atomic state at $t=0.048$ for (a) the counter-intuitive setup and (b) the intuitive one. The current in the middle wire is $I_M=0.07$ A in both cases.}
  \label{fig:Pcolor}
\end{figure}

While Fig.~\ref{fig:Populations} only gives an indication of the ongoing process as a function of time, the presence of the CTAP process for the counter-intuitively arranged wires can also be inferred from looking at the atomic probability distribution in real space. For this we show in Fig.~\ref{fig:Pcolor}  the density of the atomic state in the $x$ and $z$ plane at $t=0.048$ s integrated over the $y$-direction. At this time the atomic wavepacket is in the region where the tunnelling interaction between all three waveguides is large and clear differences between the two situations are visible. Fig.~\ref{fig:Pcolor} (a) shows the counter-intuitive situation where the wavepacket can be seen to follows the dark state with only a negligible population component in the middle waveguide. In contrast, Fig.~\ref{fig:Pcolor} (b) shows the intuitive setup, in which the population is distributed between all three waveguides and clear signatures of Rabi oscillations due to the direct tunnelling are clearly visible.

\begin{figure}[tbh]
  \includegraphics[width=\linewidth]{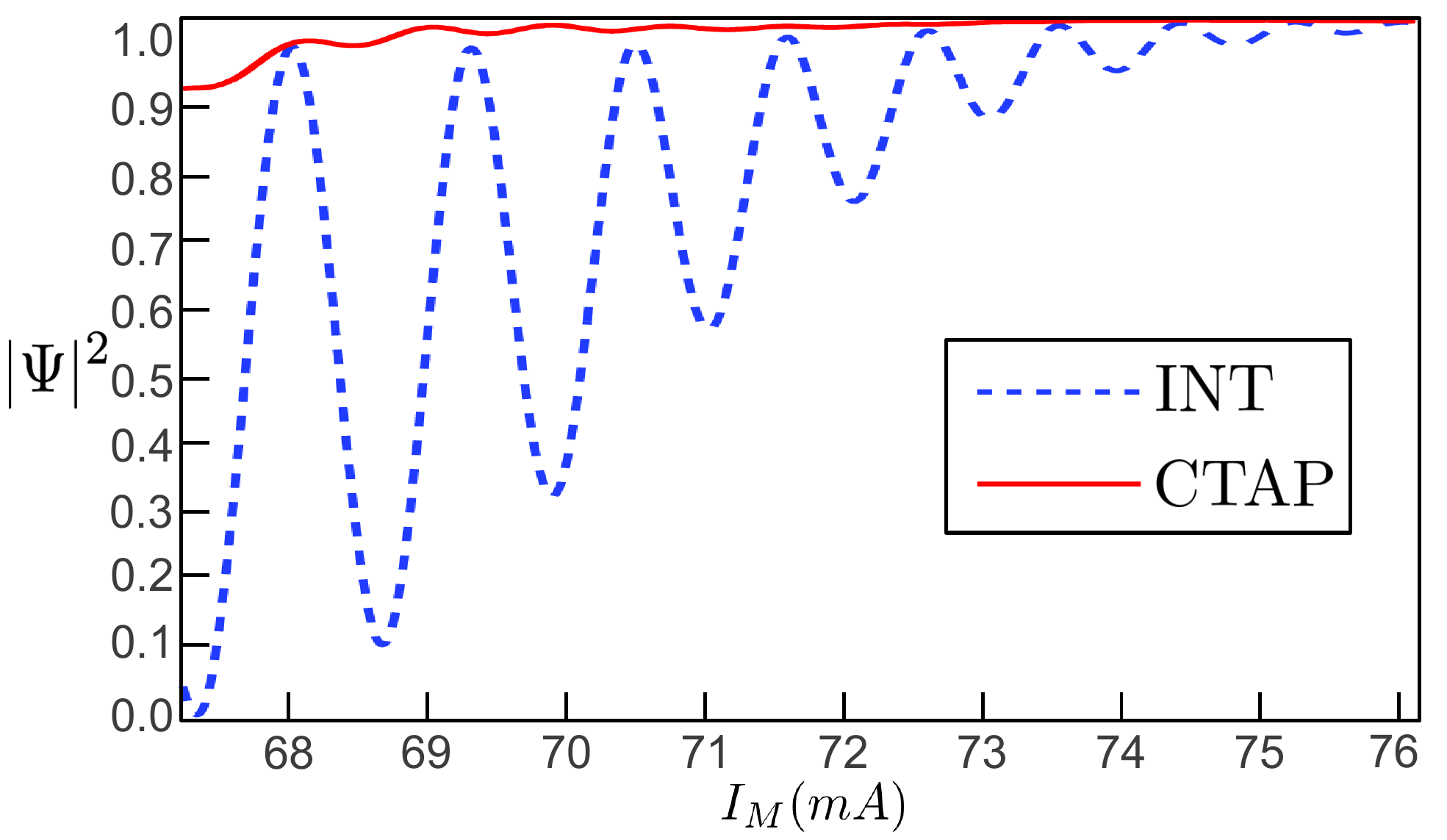}
  \caption{(Color online) The final population in the  target waveguide for both the CTAP (red solid line) and intuitive (blue dashed line) processes, for values of $I_M=0.0672$A to $I_M=0.0761$ A in steps of $0.001$ A.}
  \label{fig:Current}
\end{figure}

It is exactly these Rabi oscillations in the intuitive process that lead to the time-dependence of the final population in each waveguide and therefore a strong dependence of the outcome on small changes in the system parameters. This can be seen when examining Fig.~\ref{fig:Current}, where we show the final population in the right hand side waveguide as a function of the current in the middle wire $I_M$. For the intuitive process (blue line), the final population varies significantly with changing $I_M$, whereas the counter-intuitive setup (red line) is very robust to these changes, with the fidelity of population transfer never dropping below $0.98$. This is another indication that the transfer is due to CTAP. 

From Fig.~\ref{fig:Current} it is also clear that, while there are large oscillations in the fidelity of the intuitive process, there is an upward trend in the fidelity of the process towards unity as the current in the middle wire increases. However, at these higher values of the middle wire current, the waveguides are no longer resonant at all times and one would expect that neither the CTAP nor the intuitive processes would lead to high fidelity transfer. Nevertheless, the simulations show that this is not the case.

We conjecture that in the regime of larger currents in the middle wire the population transfer is due to  Stark-shift-chirped rapid-adiabatic-passage (SCRAP) \cite{Bergmann:05}. In this process a time-depend shift of the energy of the intermediate state in the traditional three-level arrangement allows for high-fidelity population transfer  between two states, independent of being in the intuitive or counter-intuitive situation. A translation of this to the spatial realm is straightforward: the approach and retreat of the outer wires from the middle one shift the energy of the central waveguide in a spatially dependent manner. This effect is the topic of a future investigation. 
\section{Conclusions}
\label{sec:Conclusions}

We have performed fully three dimensional simulations of an experimentally realistic waveguide system on an atom chip, where the arrangements of the wires produce spatial dependent tunnel-couplings between the waveguides. These simulations were done by implementing the CUFFT library provided by Nvidia, which made this problem numerically tractable on a desktop computer. 

Using a simple method for controlling the resonance as the waveguides are brought close together, we have demonstrated that  a counter-intuitive approach of the outer wires to the middle allows to observe high fidelity and robust transfer between the wires due to CTAP. In contrast, for intuitively coupled waveguides, where direct tunnelling between them is allowed to occur, significant Rabi oscillations between all guides exist. This makes the transfer process highly sensitive to the system parameters. While a large number of theoretical works on CTAP exist, the analysis presented offers a direct way for experimental observation and confirmation of the effect.

Finally, we have also seen a first indication that waveguide systems might be natural systems to observe the  SCRAP protocol and a detailed investigation will be the topic of a future work. While we have used the numerical methods described here to perform three dimensional simulations, they can actually be used in any number of dimensions, where they still offer large performance gains over standard CPU approaches. 

\section{Acknowledgements} 
The authors would like to thank Peter Kr\"uger and Thomas Fernholz for valuable discussions and ICHEC for use of their computing resources for the MPI-enabled code. This work was supported by Science Foundation Ireland under project number 10/IN.1/I2979.

\end{document}